\documentclass[aps,prd,showpacs,twocolumn,floatfix,showkeys]{revtex4}
\usepackage{graphicx}% Include figure files
\usepackage{dcolumn}% Align table columns on decimal point
\usepackage{bm}% bold math
\begin{document}

\title{\Large
CHARM2010: Theory Summary}

\author{
T.Barnes$^{a,b}$\footnote{Address from 3 Jan. 2011: U.S. Department of Energy, 
Office of Nuclear Physics.}}

\affiliation{
$^a$Physics Division, Oak Ridge National Laboratory,
Oak Ridge, TN 37831-6373, USA\\
$^b$Department of Physics and Astronomy, University of Tennessee,
Knoxville, TN 37996-1200, USA}

\date{\today}

\begin{abstract}
This invited summary gives some concluding remarks regarding theoretical aspects of the research presented at Charm2010.
I will specialize to the role of theory and the relative reach of theory and experiment in three of the major areas
of charm physics address at this conference, specifically 1) charm production, 2) charm weak decays, and 3) charm hadron
spectroscopy. After a discussion of the status of progress on representative topics in each of these areas I will
conclude with a previously unrelated Feynman story from a conference in the early days of charm.
\end{abstract}


\maketitle

\section{Introduction}
Charm2010 comprised only plenary presentations. Approximately 30 of these were theory, or largely addressed
theoretical issues. Most of these presentations were ``high density'', with a mean of something like 30 slides / talk.
In this 25 minute presentation I am therefore attempting to summarize about 900 slides,
which allows $\sim 1.5$ seconds / slide. In view of the short time available I will cite only a few highlights
from three principal areas of charm physics that were addressed in the conference.

Of course in a conference summary talk the one thing you should not do is to simply summarize the conference,
since the audience has just attended it as well, is presumably saturated with the material, and includes experts
who are more knowledgeable about most of the topics that were addressed. You should therefore also attempt to entertain.
For this reason, after citing some high points of theory at Charm2010, I will proceed to a previously unreported
Feynman story from the early days of charm, {\it ca.} 1973-4. My outline is thus
1. Introduction, 2. Charm2010 Theory: Executive Summary Summary, and 3. Feynman Story.

\section{Charm2010 Theory: Executive Summary Summary}

\subsection{Topics}

Although a broad range of topics in charm physics was discussed at Charm2010, one could discern three main areas of
research, which are
\vskip 0.3cm
\noindent
1) charm production

\noindent
2) charm decays

\noindent
3) charm spectroscopy
\vskip 0.3cm
We shall proceed through these sequentially, and discuss the theoretical status and recent developments, including
a summary of speakers and a few highlights in each area.

\subsection{Charm Production}

The status of calculations of charm and charmonium production was a major topic at this conference; the principal speakers
in this area and their contributions are as follows:
\noindent
J.Soto,\cite{Soto} {\it Overview of Charmonium decay and production from NRQCD};
G.Bodwin,\cite{Bodwin} {\it NRQCD factorization and quarkonium production at hadron-hadron and
ep colliders};
P.Pakhlov,\cite{Pakhlov}
{\it Double Charmonium production from B factories};
J.X.Wang,\cite{Wang}
{\it QCD correction to $J/\psi$ production at different energy scales}.
\vskip 0.3cm

All these closely related presentations were concerned with the theory of ``hard'' charm production, using the
NRQCD formalism that relates the full production amplitudes to the product of a pQCD part times and a nonperturbative
hadronic matrix element $\langle O \rangle$. The pQCD calculation involves a double expansion in $\alpha_s$ and $v_Q$,
and there are concerns about the convergence of these expansions. This approach has many applications to both
$c$ and $c\bar c$ production reactions and decays.

The application of this approach to the relatively recently exploited double charmonium production process
($e^+e^- \to (c\bar c)(c\bar c)$, as discussed in detail at Charm2010 by Pakhlov,\cite{Pakhlov}
provides an interesting case study
of NRQCD calculations of charmonium production. In this case three phases were apparent,
\vskip 0.3cm
\noindent
1) An initial leading order (LO) calculation of double charmonium production was carried out.

\vskip 0.3cm
\noindent
2) Comparison with experiment showed disagreement by an order of magnitude.

\vskip 0.3cm
\noindent
3) Theorists in response next carried out the much more difficult NLO and NNLO calculations, and found results
that were comparable to experiment.

\vskip 0.3cm
Although the final result is reassuring, it is clear that experiment plays a crucial role;
the goal in this field at present appears to be to reproduce experiment by improving calculations for
a few benchmark reaction predictions, rather than to make many new predictions.
(Theory is pursuing Experiment.) In view of the great technical
difficulty of the higher order calculations, this situation in understandable.

\begin{figure}[ht]
\vskip -0mm
\includegraphics[width=0.8\linewidth]{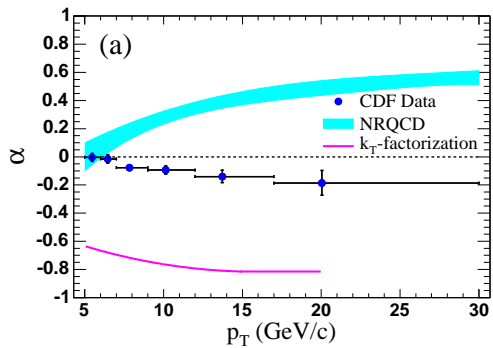}
\vskip -0cm
\caption{CDF Run II data, showing disagreement with previous low order NRQCD pQCD
predictions for polarized $J/\psi$ production.}
\label{CDF_RunII}
\end{figure}

It was quite interesting to learn that some familiar conclusions of pQCD calculations are quite sensitive
higher order corrections when compared to new experimental results. A dramatic example discussed at
this meeting was the new measurement of $J/\psi$ polarization at CDF Run II
(Fig.\ref{CDF_RunII}) (from Abulencia {\it et al.} \cite{Abulencia}, discussed here by Bodwin.\cite{Bodwin})

%%%%%%%%%%%%%%%%%%%%%%%%%%%%%%%%%%%%%%%%%%
%%%%%%%%%%%%%%%%%%%%%%%%%%%%%%%%%%%%%%%%%%

%%%%%%%%%%%%%%%%%%%%%%%%%%%%%%%%%%%%%%%%%%
\begin{figure}[ht]
\vskip -0mm
\includegraphics[width=0.8\linewidth]{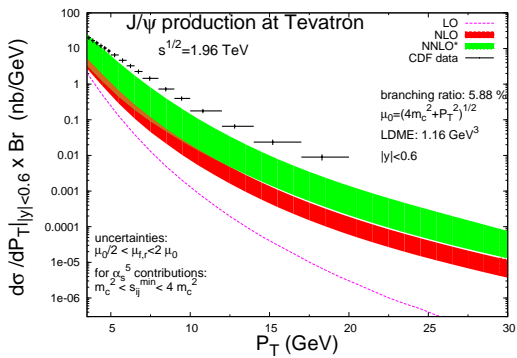}
\vskip -0cm
\caption{NRQCD calculation of the color-singlet J/Psi-production differential cross section at the
Tevatron. Inclusion of higher-order effects considerably increases the estimated color-singlet
contribution, which may now dominate the color-octet component.}
\label{Artoisenet}
\end{figure}
%%%%%%%%%%%%%%%%%%%%%%%%%%%%%%%%%%%%%%%%%%
Including this data in pQCD
fits has changed the favored mix of color octet to color singlet production amplitudes.
Previously, color-octet dominance was thought to be well established. The color-singlet amplitudes
have been found to have large high order corrections, and when compared to the new CDF data, it is found that
color-singlet production may actually be dominant (Fig.\ref{Artoisenet}).

Another new development in this area of theory is in the values used for the nonperturbative
hadronic matrix elements $\{\langle O \rangle\}$. Previously these have been estimated from models
or treated as free parameters in fits to data. These quantities may instead be calculated directly
using lattice QCD, which will be very important in removing a major source of uncertainty from these
calculations.

\subsection{Charm Decays}

In addition to learning about properties of QCD through charm production and
charm hadron spectroscopy, charm physics can also be used to quantify and test aspects of the
standard model through studies of charm weak decays. This was also a major topic at Charm2010,
with theoretical contributions (and some experimenters who nicely summarized theory) as follows:
Hai-Yang Cheng,\cite{Cheng}
{\it Theoretical review on Hadronic D/D$_s$ decays};
Jernej F.Kamenik,\cite{Kamenik}
{\it V$_{ub}$ and weak annihilation in inclusive semileptonic D/D$_s$ decays};
Heechang Na,\cite{Na}
{\it Recent progress on D semileptonic decays from Lattice QCD};
James Simone,\cite{Simone}
{\it Recent results on decay constant of charm mesons from LQCD};
Sebastian Descotes-Genon,\cite{Descotes-Genon}
{\it CKM fitter and the role of Charm decays};
Alexander Lenz,\cite{Lenz}
{\it SM Predictions on D$^0$ oscillations and CPV};
Gilad Perez,\cite{Perez}
{\it Theoretical review on the prospect for new physics in charm sector};
Jim Libby,\cite{Libby}
{\it The Impact of Quantum Correlations at Charm Threshold on the Determination
of CKM Angle $\gamma$};
David Asner,\cite{Asner}
{\it Charm Mixing and Strong Phases at Threshold};
Jerome Charles,\cite{Charles}
{\it CPV in (DD) pairs decay into (V1V2)(V3V4) and extraction of strong phase
at threshold}.

The charm weak decay topics addressed at Charm2010 were primarily D and D$_s$ leptonic and
semileptonic decays, and to a lesser extent D$^0$-${\bar {\rm D}}^0$ oscillations.
The goals of these studies are to establish the entries of the CKM matrix, test unitarity, and
also to search for evidence of new physics in the ``$u$'' sector.

The weak matrix elements for D and D$_s$ leptonic and
semileptonic decays (rather like pQCD charm production calculations) involve a
nonperturbative hadronic matrix element or form factor, $f_{mes}$ or $f_{mes}(Q^2)$, times a CKM matrix element,
$V_{cd}$ or $V_{cs}$.
Current practice is to determine the value of this hadronic matrix element or a parametrization
of these form factors using lattice QCD.

\begin{figure}[ht]
\vskip -0mm
\includegraphics[width=0.8\linewidth]{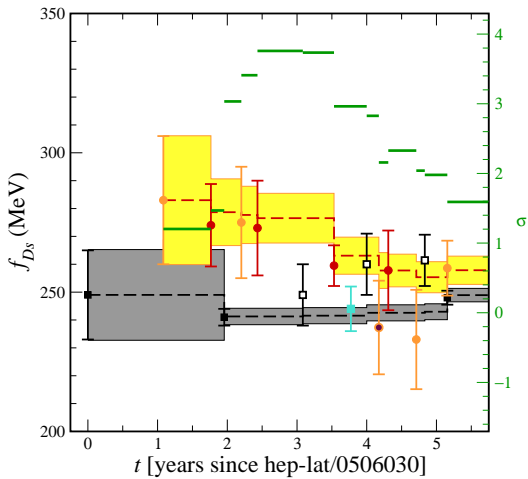}
\vskip -0cm
\caption{The recent convergence between experimental (upper bands) and theoretical LQCD (lower bands)
values for the nonleptonic weak D$_s$ coupling parameter $f_{D_s}$ (LHS scale). The discrepancy
(short horizontal line segments) in $\sigma$ references the RHS scale.}
\label{fDs}
\end{figure}

Rather remarkably there appears to be a general state of happiness in this field, with
experimental and theoretical errors comparable (or soon expected to be) at the few-$\%$ level.
An apparent 3.8$\sigma\; $discrepancy in the D$_s$ case has been resolved ``adiabatically''
during the previous three years, through increased lattice and decreased experimental estimates
of $f_{Ds}$.
(See Fig.\ref{fDs}.\cite{Descotes-Genon,Kronfeld})

%%%%%%%%%%%%%%%%%%%%%%%%%%%%%%%%%%%%%%%%%%
%%%%%%%%%%%%%%%%%%%%%%%%%%%%%%%%%%%%%%%%%%

%%%%%%%%%%%%%%%%%%%%%%%%%%%%%%%%%%%%%%%%%%
\begin{figure}[ht]
\vskip -0mm
\includegraphics[width=0.8\linewidth]{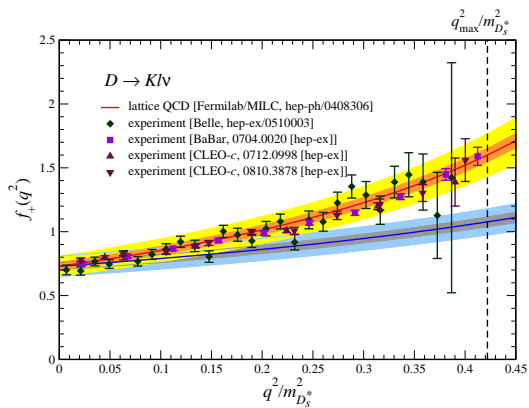}
\vskip -0cm
\caption{A recent comparison$^{18}$ between theoretical LQCD (upper line) and experimental results for
the D$ \to $K$ l \nu $ semileptonic weak decay form factor.}
\label{slff_DK}
\end{figure}
%%%%%%%%%%%%%%%%%%%%%%%%%%%%%%%%%%%%%%%%%%

The semileptonic decays can similarly be used to estimate $V_{cd}$ or $V_{cs}$, and
lead to consistent values of these CKM matrix element entries. A recent D$ \to $K$ l \nu $
example similar to those presented at Charm2010 is shown in 
Fig.\ref{slff_DK}.\cite{Bernard:2009ke})

In contrast, D$^0$-${\bar {\rm D}}^0$ oscillations involve a strong relative phase between the amplitudes for
$\imath \to f$ and $\bar \imath \to f$ processes, which cannot yet be calculated reliably in general.
This strong can be either estimated or controlled by clever manipulation of Dalitz plots, or avoided
completely through studies of ``golden mode'' transitions that have zero relative strong phase.

There was a single contribution to charm decays that addressed strong decays (rather than weak),
which was concerned with strong decays of charmonia to nucleon, antinucleon and light meson final states;
T.Barnes,\cite{Barnes}
{\it Meson Emission Model of $\Psi \to N\bar N m$ in Charmonium Strong Decays}.
This contribution noted that experimental results for these decay partial widths and Dalitz plot
event densities (some of which have already been measured at BES and CLEO) could be used to estimate
cross sections for associated charmonium production ($p \bar p \to \Psi m$) at PANDA, and that these
decays can also be used to estimate $NNm$ couplings, which are crucial inputs for meson exchange
models of the nuclear force. $J/\psi \to p \bar p \omega$ was suggested as an interesting test case for
BES.

\subsection{Charm Spectroscopy}

Finally, theoretical aspects of charm hadron spectroscopy (in both open- and closed-charm systems)
was the third major theme of Charm2010. Theory contributions in this area are as follows:
J.Vijande,\cite{Vijande}
{\it On the nature of the X(3872)};
C.DeTar,\cite{DeTar}
{\it Charmonium and charm spectroscopy from Lattice QCD};
M.Nielsen,\cite{Nielsen}
{\it Theoretical review on exotics charmonium};
E.Oset,\cite{Oset}
{\it The X(3872) and other X,Y,Z resonances as hidden charm meson-meson
molecules};
Ulf-G.Meissner,\cite{Meissner}
{\it Open charm and charmonium states from EFTs};
X.Liu,\cite{LiuX}
{\it Theoretical review on excited D*/D$_s$* mesons};
R.Molina,\cite{Molina}
{\it A new interpretation for the D$_{s2}$*(2573) and the prediction of novel
exotic charmed mesons};
J.-M.Richard,\cite{Richard}
{\it Baryon resonances};
A.Valcare,\cite{Valcare}
{\it Exotic charmed four-quark mesons: molecules versus compact states};
P.Gonzalez,\cite{Gonzalez}
{\it A precise quark model description of charmonium spectrum};
C.Liu,\cite{LiuC}
{\it Low energy D$^{0*}$ D$_1^0$ scattering and Z(4430) from LQCD};
Q.Zhao,\cite{Zhao}
{\it Effect of charmed meson loops on charmonium transitions}.

The large number of contributions in spectroscopy is an indication of the level of research activity in
charm spectroscopy in recent years; due in part to the B factories,
this has recently been a very active field experimentally.
The topics addressed at Charm2010 included
attempts to understand the now very rich spectrum of $c$ and $c\bar c$ hadrons using models
or numerical QCD; specific topics included $c\bar c$ potential models, LQCD (lattice QCD),
QCD sum rules, hadron-level lagrangians (EFTs), heavy-quark baryons,
coupled channel effects, LQCD simulations of scattering (interhadron forces), and meson loop effects.

\begin{figure}[ht]
\vskip -0mm
\includegraphics[width=0.8\linewidth]{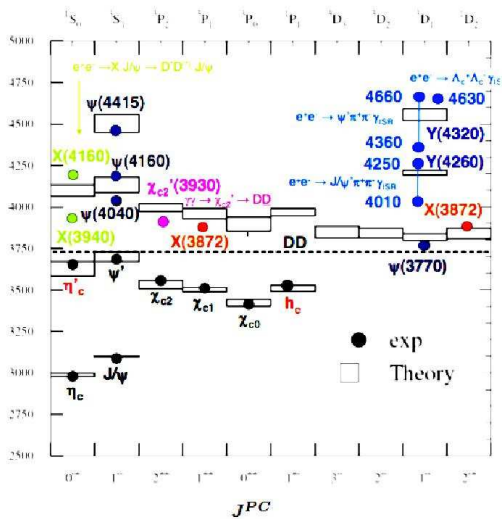}
\vskip -0cm
\caption{The expected spectrum of charmonia, compared to some of the recent experimental states.}
\label{Nielsen_spectrum}
\end{figure}

\begin{figure}[ht]
\vskip -0mm
\includegraphics[width=0.8\linewidth]{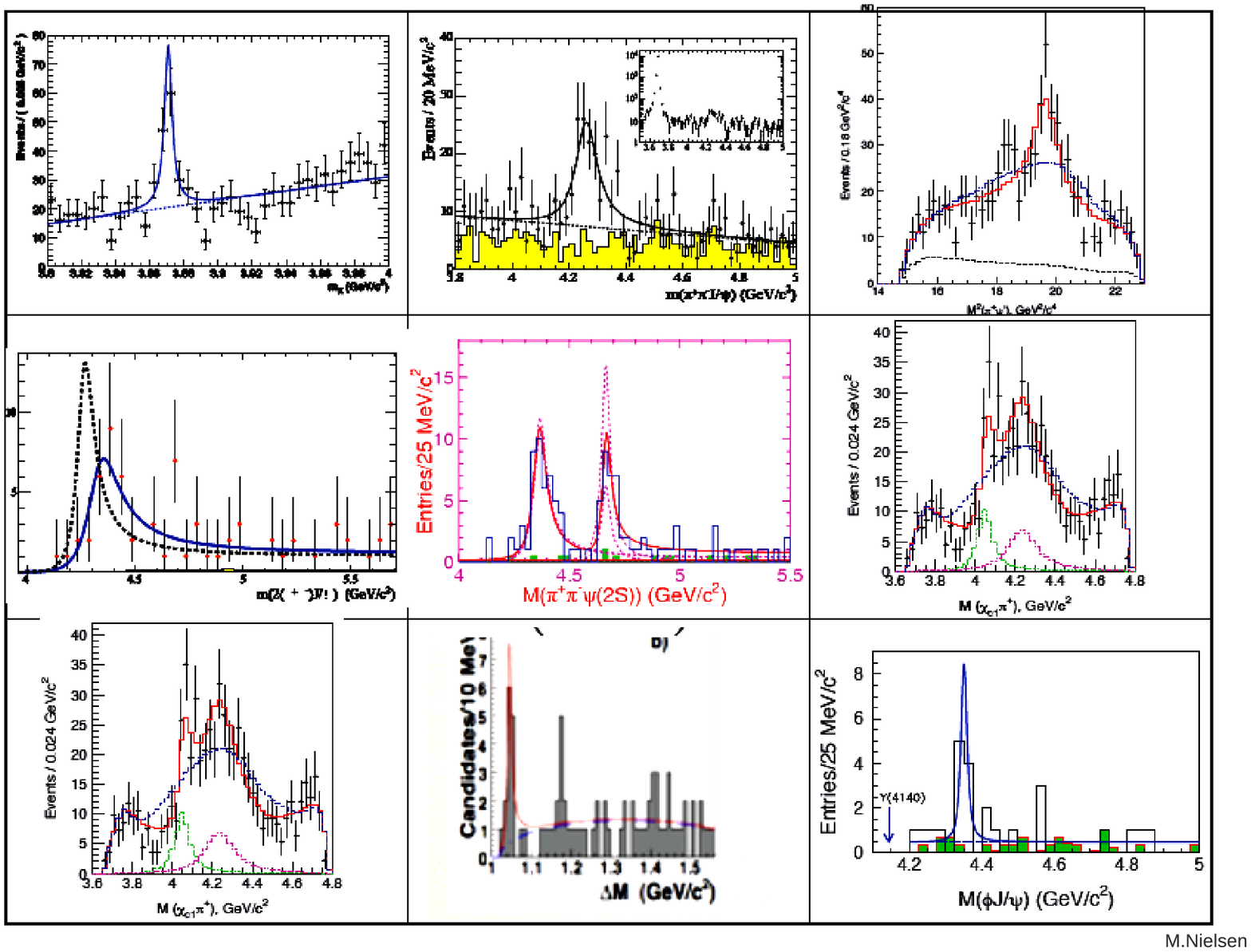}
\vskip -0cm
\caption{A collage of data on nine of the recently reported charm sector meson resonance candidates,
as discussed by Nielsen. The states are (from top left to bottom right) the
X(3872), Y(4260), Z+(4430); Y(4360), Y(4660), Z1*(4050); Z2+(4250), Y(4140) and X(4350).}
\label{MN_3by3matrix}
\end{figure}

The great increase in theoretical effort, and the perhaps unsettling lack of general conclusions
regarding the interpretation of most of the new experimental discoveries, illustrates the richness
and chaotic recent growth of charm spectroscopy. Until recently it was generally believed that
potential models gave an accurate description of the spectrum of charm hadrons. However, with the
discovery of the D$^{*0}$(2317) in 2003, some $\approx 150$ MeV below the predicted mass, it
became clear that potential models had missed some crucial aspects of the physics. We now have
a large number of new states in these sectors (see Nielsen\cite{Nielsen} for a Charm2010 talk
on this subject), and only {\it one} of the new states has a compelling assignment as a conventional
quarkonium state (the Z(3930), seen in $\gamma\gamma$, which meets expectations for a radially excited
2$^3$P$_2$ $c\bar c$ state).
Two slides from Nielsen's Charm2010 review talk\cite{Nielsen} showing a $c\bar c$ level diagram
versus new state masses and the data for some of the states are shown in Figs.\ref{Nielsen_spectrum},\ref{MN_3by3matrix}.

%%%%%%%%%%%%%%%%%%%%%%%%%%%%%%%%%%%%%%%%%%
%%%%%%%%%%%%%%%%%%%%%%%%%%%%%%%%%%%%%%%%%%

%%%%%%%%%%%%%%%%%%%%%%%%%%%%%%%%%%%%%%%%%%
%%%%%%%%%%%%%%%%%%%%%%%%%%%%%%%%%%%%%%%%%%

Of the many other new states, the X(3872) attracted the most attention
at Charm2010, since it appears plausible as a candidate for a remarkable, weakly-bound
D$^{*0}$D$^0$ meson molecule (Hermitian conjugates are implicit).
The strongly isospin-violating decays of the X(3872) were anticipated for a neutral-D and
neutral-D$^*$ molecule candidate, and this state (an S-wave $1^{++}$ combination)
was predicted as a result of the known
one pion exchange couplings.

Determining the assignments of the many new states as conventional quarkonia, hybrids, mesonic molecules,
or linear combinations of these nominal basis states will be the task of future theoretical studies.
It was noted in some presentations that $q^2{\bar q}^2$ “tetraquark” clusters are dubious
assignments for these new states, so there has been {\it some} progress in theory; like pentaquarks,
these multiquark clusters probably fall apart unless below their 2-hadron thresholds.
Presumably some of the basic physics (such as the hybrid spectrum and preferred hybrid strong decay modes,
and the interhadron forces that lead to molecular states) can be established through LQCD studies, which can
then be implemented in more detailed models. Clearly there is much work ahead for theorists in charm spectroscopy,
which has proven to be an amazingly rich field.

\section{Feynman Story}

\noindent
(follows)

\vfill\eject

Physicists attend a large number of conferences, perhaps as many as several hundred in the course of a career.
Do you recall your first conference? Mine was in Dec.~1973, the {\it Irvine Conference on Lepton-Induced Reactions},
which took place when I was a relatively new graduate student at Caltech. We high energy theory graduate students
learned that Feynman was planning to attend the meeting, and since UC Irvine was about 30~miles from Caltech,
he had volunteered to drive
any graduate students who were interested. Several of us asked for places in his car, and were surprised and puzzled
to learn that our request was successful! (What about the competition from more senior physicists?) This was great;
we would have Feynman to ourselves to discuss physics with for at least several hours, during the drive from Pasadena
to Irvine and back.

So, I and the other graduate students arrived at Lauritsen Lab. very early in the morning, and met Feynman
in the parking lot.
He had brought a rather unimpressive old car, which we climbed into; I recall being safely hidden in the left rear seat.
As we started it quickly became apparent why we had had no trouble getting places in Feynman's car - he was a really
bad driver. REALLY bad. (Perhaps you have read this in histories that mention Feynman? I can confirm it.) He almost
ran into another car pulling out of the Lauritsen lot onto Calif. Blvd., ``{Whazza maddah with that guy?}''
and proceeded down the road to the Pasadena freeway at highly variable speeds, not maintaining his lane well,
and already talking about physics. This was very stressful; you were trying to think about and rationally talk
about physics with the world's leading theorist, who was a very `in your face' speaker who spoke quickly with
an intense, insistent Brooklyn accent, while fearing for your life ... and Feynman REALLY wanted to talk
physics with us. In addition
to his natural love of physics, I think this was in part because he was conference summary speaker
at Irvine, and wanted to try out some new ideas on us.

While driving down the crowded freeway at high speeds, Feynman would suddenly have an exciting idea, and
since three of us were in the back seat he would turn completely around and enthusiastically explain his
thoughts to us. Occasionally this would involve drifting out of his lane, with an angry response from other
drivers. We mere graduate students of course could not tell Feynman to turn back around and
concentrate on his driving, and save the physics for later.
While continuing down the freeway in this state of moderate panic, I noticed an interesting correlation:
I could predict when Feynman would suddenly turn around and face us. While going down the freeway
his speedometer normally read something like 50-60 mph, {\it albeit} with considerable random
modulation. However it would occasionally rapidly climb to about 70 mph. This was an indication that
Feynman had an exciting new idea. After a short time at this higher speed, Feynman would suddenly turn
to the back seat and start lecturing to us. At least this unsettling maneuver was predictable!

One of the interesting new results that Feynman was excited about, which was to be reported at the conference
by B.Richter, concerned the experimental cross section for $e^+e^- \to hadrons$ observed at the highest accessible
energies ($E_{cm} \approx 3$~GeV) at SLAC. Feynman told us that there was an increase seen in this cross section,
and of course people had been speculating that this might be associated with production of the anticipated
but as yet unobserved charm quark. Feynman said that he had thought of a more conventional explanation
at breakfast, and he wanted our opinion. Perhaps instead of charm production, this increase could instead
be due to a two-photon process, $e^+e^- \to \gamma\gamma \to \rho^0 \rho^0 \to hadrons$, in which the
photons converted to rho mesons by vector dominance, followed by a strong final-state interaction.
``{Whadda you guys think about that?}'' We looked at each
other in panic; Feynman wanted us to quickly assess this new process he had suggested, while ignoring the
very real prospect of being in a fatal car crash in the immediate future.
I visualized the Feynman diagrams in question, and since I was too sleep-deprived to stop myself I quickly
blurted out ``{Too many powers of alpha.}'' Back came ``{Whadda ya MEAN too many powers of alpha?}''
I started slowly describing one of the Feynman diagrams and the powers of $e$ in each place, and Feynman immediately
started talking quickly about ``{lessee alpha alpha ... alpha$^4$}'', and then
``{Ya know, you're right!}'' He then wheeled completely around to stare intensely at me, the graduate student
who had dared correct him, and said with a big grin ``{Too many powers of alpha, HUH?}''
Since Feynman did not mention this idea during his conference summary, I may have been successful in directing
him away from it. I later learned that this two-photon process WAS considered seriously by some other
theorists; perhaps Feynman found the na\"ive graduate student objection more convincing than arguments about
large rho-rho strong cross sections?

Another thing I learned during this conference was about conference summary talks. We learned on the last day that
Feynman was in a trailer somewhere on the Irvine campus preparing his summary, which surprised me because
the lectures were still taking
place. How could you summarize a conference without attending the talks? The answer became clear during Feynman's
talk; I don't recall much about it except that it seemed to have little to do with the details of the conference.
This was a disappointment, since I had hoped that all would be made clear during his summary.
(I also recall someone offering to take over the lectern from Feynman and give his own mini-talk,
which Feynman declined on the grounds that it was probably wrong.)
I suppose the lesson here is that it may actually be inappropriate to summarize a conference
during your conference summary talk, except in a very general way, and if possible you should entertain the audience
in the process. I hope I have succeeded in this endeavor at Charm2010, almost 37 years after Irvine,
at which we are still excitedly discussing the physics of charm.

\vfill\eject

\section{Acknowledgments}

I am happy to acknowledge the kind invitation of Hai-Bo Li, Yifang Wang, and the organizers
to present this Charm2010 Theory Summary talk at Beijing.
I would also like to thank G.Bodwin, A.Kronfeld, M.Nielsen, R.Poling and J.Simone
for assistance with figures.
This research was supported in part by the U.S. Department of Energy under contract
DE-AC05-00OR22725 at Oak Ridge National Laboratory.
The support of the Department of Physics and Astronomy
of the University of Tennessee is also gratefully acknowledged.

\end{document}